\documentclass{bmcart}

\usepackage{amsthm,amsmath}
\RequirePackage[numbers]{natbib}
\usepackage[utf8]{inputenc} 
\usepackage{graphicx,url,bm,comment}
\setcitestyle{numbers}

\usepackage{tabularx}

\startlocaldefs
\newcommand{\btheta}{\bm{\theta}}

\usepackage{fancyvrb}

\DefineVerbatimEnvironment{Highlighting}{Verbatim}{commandchars=\\\{\}}
\usepackage{framed}
\definecolor{shadecolor}{RGB}{248,248,248}
\newenvironment{Shaded}{\begin{snugshade}}{\end{snugshade}}

\newcommand{\AttributeTok}[1]{\textcolor[rgb]{0.77,0.63,0.00}{#1}}

\newcommand{\DecValTok}[1]{\textcolor[rgb]{0.00,0.00,0.81}{#1}}

\newcommand{\FloatTok}[1]{\textcolor[rgb]{0.00,0.00,0.81}{#1}}
\newcommand{\FunctionTok}[1]{\textcolor[rgb]{0.00,0.00,0.00}{#1}}

\newcommand{\NormalTok}[1]{#1}

\newcommand{\OtherTok}[1]{\textcolor[rgb]{0.56,0.35,0.01}{#1}}

\newcommand{\SpecialCharTok}[1]{\textcolor[rgb]{0.00,0.00,0.00}{#1}}

\newcommand{\StringTok}[1]{\textcolor[rgb]{0.31,0.60,0.02}{#1}}

\endlocaldefs

\begin{document}

\begin{frontmatter}

\begin{fmbox}
\dochead{Research}


\title{survextrap: a package for flexible and transparent survival extrapolation}


\author[
  addressref={bsu},  
  email={chris.jackson@mrc-bsu.cam.ac.uk}   
]{\inits{C.H.}\fnm{Christopher H.} \snm{Jackson}}


\address[id=bsu]{
  \orgdiv{MRC Biostatistics Unit},             
  \orgname{University of Cambridge},          
  \city{Cambridge},                              
  \cny{UK}                                    
}



\end{fmbox}


\begin{abstractbox}

\begin{abstract} 
  \parttitle{Background} Health policy decisions are often informed by estimates of long-term survival based primarily on short-term data.  A range of methods are available to include longer-term information, but there has previously been no comprehensive and accessible tool for implementing these.
  
\parttitle{Results}
This paper introduces a novel model and software package for parametric survival modelling of individual-level, right-censored data, optionally combined with summary survival data on one or more time periods.  It could be used to estimate long-term survival based on short-term data from a clinical trial, combined with longer-term disease registry or population data, or elicited judgements.  All data sources are represented jointly in a Bayesian model.  The hazard is modelled as an M-spline function, which can represent potential changes in the hazard trajectory at any time.  Through Bayesian estimation, the model automatically adapts to fit the available data, and acknowledges uncertainty where the data are weak.  Therefore long-term estimates are only confident if there are strong long-term data, and inferences do not rely on extrapolating parametric functions learned from short-term data.  The effects of treatment or other explanatory variables can be estimated through proportional hazards or with a flexible non-proportional hazards model.  Some commonly-used mechanisms for survival can also be assumed: cure models, additive hazards models with known background mortality, and models where the effect of a treatment wanes over time.  All of these features are provided for the first time in an R package, \texttt{survextrap}, in which models can be fitted using standard R survival modelling syntax.  This paper explains the model, and demonstrates the use of the package to fit a range of models to common forms of survival data used in health technology assessments.

\parttitle{Conclusions}
This paper has provided a tool that makes comprehensive and principled methods for survival extrapolation easily usable.
\end{abstract}




\end{abstractbox}
%

\end{frontmatter}

\startlocaldefs
\newcommand{\bx}{\mathbf{x}}
\newcommand{\x}{\mathbf{x}}
\endlocaldefs




\section{Background}

Health policy decisions are often informed by censored survival data with limited follow-up, such as clinical trial data.   However, since decisions may have long-term consequences, the policy-maker is typically interested in the expected survival over the long term.  This can be difficult to estimate when the main source of data is short-term.  This task is generally referred to as ``extrapolation" \citep[e.g.][]{latimer2022extrapolation}.  While this word may imply a naive assumption that short-term trends will continue in the long term,  it is now widely acknowledged that making reliable decisions requires combining evidence and judgements about both the short term and the long term.  Many approaches have been suggested for combining short-term and long-term data for survival extrapolation.   For reviews of these methods, see \citet{bullement2023systematic,jackson2017extrapolating}, and for a broader review of flexible models for survival analysis in health technology assessments, see \citet{nicedsu:flexsurv}.

An overview of these approaches is now given, structured around four desirable characteristics: (a) to allow all relevant information to be included, (b) to fit the data well, (c) to allow the resulting uncertainty to be quantified, (d) to be easy to implement.

\paragraph{Including all relevant information}
As well as the short-term trial data, there may be registry data on people with the disease of interest, or national statistics describing mortality of the general population.  Long-term survival can then be estimated by building a model to jointly describe all data sources \citep{demiris2011survival,benaglia2015survival,guyot2017extrapolation}, under assumptions on how the data are related, for example, proportional hazards between populations.  Another common approach is based on partitioning time into different intervals in which the hazard is informed by different data \citep[e.g.][]{che2022blended}.  Elicited expert judgements, ideally about interpretable quantities such as survival probabilities, can also be included in survival extrapolation, either to directly provide long-term survival estimates \citep{cope2019integrating} or through formal Bayesian combinations with data \citep{cooney2023direct}. 

As well as including all relevant \emph{data}, models should also be designed to represent any \emph{substantive knowledge} about how risks vary with time and between people.  One common assumption involves distinguishing different causes of death in the model (e.g. the cause of interest and background causes) through additive hazards \citep{van2021comparison,nicedsu:flexsurv,nelson2007flexible}.    Another commonly-modelled mechanism is the notion of ``cure" \citep[e.g][]{boag1949maximum,federico2022heterogeneity,chaudhary2022use}, where some survivors are eventually assumed to have zero or negligible risk of some type of event in the long term.   A particularly important quantity in healthcare decision models is the relative effect of a new treatment on survival, which is generally unknowable beyond the horizon of its trial, and modellers are often advised to consider different mechanisms for how this effect might change \citep{nice:guide}.

\paragraph{Faithfully representing observed data}
A relatively easy part of this task is to estimate short-term risks from the short-term data.  There is a vast range of statistical methods available for building, selecting or averaging over parametric survival models~\citep{nicedsu:flexsurv,kearns2021comparing}.  This is important to do well, since the expected survival in the long term is a function of both short-term and long-term hazards.  However, short-term fit is a weak guide to long-term plausibility.  Extrapolations of fitted models outside the data are heavily influenced by the choice of parametric form, therefore are unreliable if this is not informed by a plausible mechanism.   The most common survival models (e.g. Weibull and log-logistic) are generally chosen for their mathematical convenience and availability in software, rather than their biological plausibility~\citep{bagust:beale}.    Flexible parametric survival models, e.g. spline models~\citep{royston:parmar} are designed to adapt their shape to fit data arbitrarily well.  These allow the shape of the hazard function to change at any time, hence can adapt to fit combinations of short-term data and long-term data \citep{guyot2017extrapolation,vickers2019evaluation}.   Since the shapes of fitted spline models are driven by data, rather than knowledge about the mechanism, caution is advised when using them for extrapolation \citep{nicedsu:flexsurv,kearns2021comparing}. 

\paragraph{Expressing uncertainty about survival}
An appreciation of uncertainty is important in healthcare decision-making \citep{briggs2012model}. Representing parameter uncertainty probabilistically (the Bayesian perspective) has various advantages \citep{Spiegelhalteretal:2004} --- one advantage is the ease with which multiple sources of evidence can be modelled jointly to enhance information  about quantities of interest.   This approach, sometimes called ``multiparameter evidence synthesis", has been used for survival extrapolation~\citep{demiris2011survival,benaglia2015survival,guyot2017extrapolation,vickers2019evaluation,kearns2021comparing,van2021comparison,chaudhary2022use}.  Another advantage of the probabilistic perspective is that it allows the expected value of further information to be calculated, e.g. for longer-term follow up of survival \citep{vervaart2022efficient,tai2021prevalence}. 

\paragraph{Ease of implementation}
For a statistical method to be useful, it should be as easy as possible to use in software.  The ideal tool would allow the decision-maker to input all available data and relevant knowledge, and convert those to interesting results.  Relevant assumptions should be made transparent, while the computer bears the burden of translating knowledge and assumptions to a mathematical form and processing the data.  Flexible survival models can easily be fitted to individual-level censored data, for example using the R packages \texttt{flexsurv}~\citep{jackson2016flexsurv} or \texttt{survHE}~\citep{baio2020survhe}, or the Stata package \texttt{stpm2}~\citep{lambert2009further}, but these do not have facilities for including ``external" data to aid extrapolation.   Bayesian evidence synthesis models have been implemented using Bayesian modelling languages, e.g. BUGS~\citep{guyot2017extrapolation} and JAGS~\citep{vickers2019evaluation}, though these require specialised statistical and programming skills.

\subsection{The \texttt{survextrap} model and package}
In this paper's view, there has been no method for survival extrapolation that satisfies all four of these desirable criteria.  For example, while the \citet{guyot2017extrapolation} model flexibly accommodates multiple sources of data, it requires specialised programming and advanced statistical expertise.  The method of \citet{che2022blended} is based on probabilistically blending a model for short-term data with a model for long-term data, however this only accommodates two sources of data, and does not address how to develop and implement each of the two models.


This paper presents a model and R package, \texttt{survextrap}, that is intended to achieve these criteria.  The model is a Bayesian evidence synthesis, that can combine right-censored individual data with any number of external data sources.  External data are supplied in a general aggregate count form that can encompass typical population or registry data, as well as elicited judgements about survival probabilities.   A penalised spline model is used that can represent hazard changes at any times.  Through a Bayesian procedure, an appropriate level of smoothness and flexibility is estimated.  The result is a posterior distribution that represents uncertainty about survival given all data and knowledge provided.   Uncertainty about potential hazard changes in times not covered by the data can also be included in this posterior, by allowing the spline function to vary smoothly in these times.  The package also implements some commonly-used mechanisms for survival extrapolation: additive hazards (relative survival) models, mixture cure models, and models where the effect of a treatment wanes over time.    A model can be fitted in \texttt{survextrap} using a single call to an R function, which follows the standard syntax for survival modelling in R, and a range of common summary outputs can be extracted with single function calls.

The model is fully described in Section~\ref{sec:model}, explaining the idea of M-splines, how they are used to model data, and their extensions to deal with explanatory variables and special mechanisms.   Section~\ref{sec:impl} introduces the \texttt{survextrap} package and points to an example of its basic use. Section~\ref{sec:demo} demonstrates how the model and package might be used in a realistic application, to an evaluation of the survival benefits of cetuximab in head and neck cancer~\citep{guyot2017extrapolation}.  A range of models are compared, each with different data sources and assumptions about how inferences outside the data are made.   The discussion (Section~\ref{sec:discussion}) gives some suggestions for further research. 

\section{Methods: statistical model}
\label{sec:model}

We suppose there is:
\begin{itemize}
\item[(a)] an individual-level dataset, with survival times that may be right-censored,
\item[(b)] optionally, also one or more aggregate external datasets, giving counts of survivors over arbitrary time periods.
\end{itemize}
The external datasets, indexed by $j=1,\ldots,J$, take the following form:
\begin{itemize}
\item out of $n_j$ people alive at a time $u_j$, with common characteristics $\x_j$,
\item $r_j$ of them survive until the time $v_j$.
\end{itemize}
This form of data might be derived from e.g. disease registries or population life tables.  It may also be derived from expert elicitation of the survival probability $p_j$ over the period $(u_j,v_j)$, using the following method.  Suppose we elicit a $Beta(a,b)$ prior for $p_j$.  We interpret this as the posterior distribution from a vague prior (Beta(0,0), say) for $p_j$ combined with data $(r_j, n_j)$, which would be a $Beta(r_j,n_j-r_j)$.  Then, by equating $a=r_j, b=n_j-r_j$, we can deduce the data $(r_j, n_j)$ that would represent knowledge equivalent to the elicited judgement. See~\citet{cooney2023direct} for a related approach.

A single model is assumed to jointly generate all sources of data.  This is defined by its hazard function
$h(t \vert \btheta, \x)$,
where $t$ is time, $\btheta$ includes all parameters, and $\x$ includes predictors (e.g. characteristics of individuals, or variables that distinguish one dataset from another).  This model will be based around a flexible function known as an M-spline~\citep{ramsay1988monotone}, as previously used by~\citet{brilleman2020bayesian} and \citet{frailtypack} for survival modelling.  M-splines have some computational advantages over other kinds of splines and flexible models, as discussed in the Appendix. 

Sections~\ref{sec:mspline}--\ref{sec:covs} describe the ingredients of this model in more detail.   This core model can then form the basis of some other specialised survival modelling mechanisms: additive hazards, cure models and treatment waning,  as described in Section~\ref{sec:mechanisms}.    The model will be fitted by Bayesian inference (Section~\ref{sec:bayes}), which produces a posterior distribution for the parameters $\btheta$.  Estimates and measures of uncertainty for long-term survival and other quantities of interest can then be deduced.


\subsection{M-spline model}
\label{sec:mspline}

In an M-spline model, the hazard $h(t)$ at time $t$ is defined by a weighted sum of \emph{basis functions}, which takes the form:

$$h(t) = \eta \sum_{i=1}^n p_i b_i(t)$$ 

\begin{figure}
  \centering
  \includegraphics{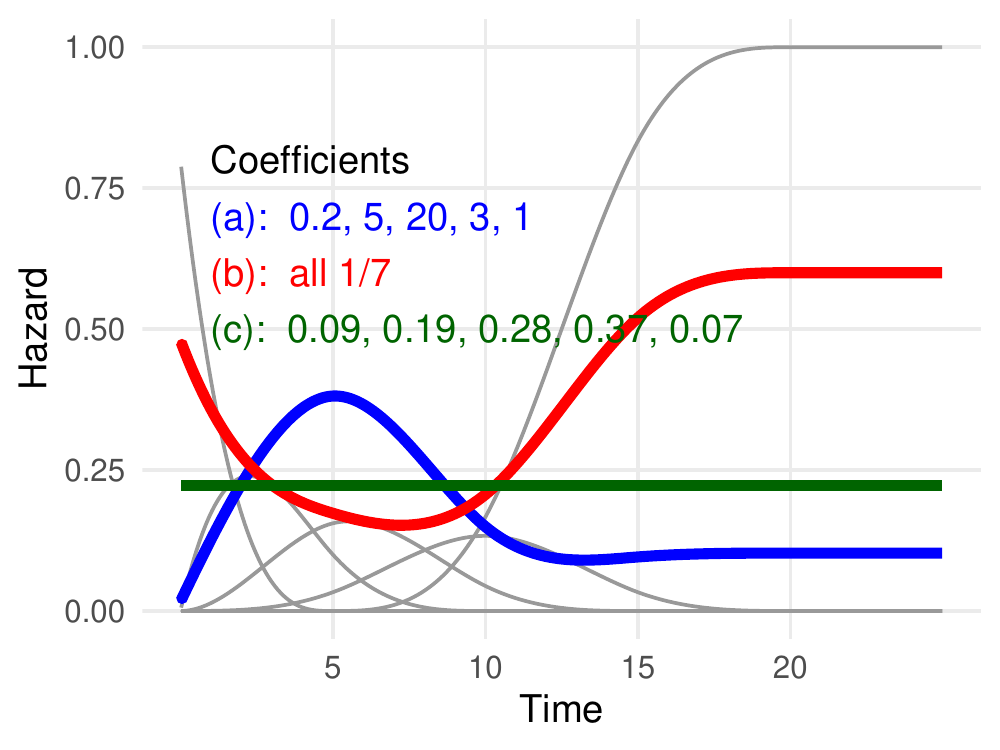}
  \caption{Three theoretical examples of M-spline hazard functions $h(t)=\sum_i p_i b_i(t)$ (coloured lines). Each is defined as a linear combination of the same 5 polynomial basis functions $b_i(t)$ (in grey), and scale $\eta=3$.  Each has different basis coefficients $p_i$, indicated in the plot. Note that the constant hazard model is not the one where all the coefficients are the same.}
  \label{fig:mspline}
\end{figure}

The \emph{scale} parameter $\eta$ is proportional to the typical level of the hazard, and the \emph{basis coefficients} $p_1,\ldots,p_n$ satisfy $\sum_i p_i = 1$.   A simple example is illustrated in Figure~\ref{fig:mspline} with $n = 5$ basis functions.   The axis of time is split into regions defined by $n-1$ ``knots'', in this example at 5, 10, 15 and 20.   Given these knots, a set of $n$ basis functions $b_i(t)$ are defined to span the range of the knots, where the first one has a peak at zero, the next $n-2$ functions have peaks inside the knots, and the final basis function is constant when $t$ exceeds the final knot.  The basis functions are polynomials (cubics by default), restricted to be positive and to ensure that $h(t)$ is a smooth function of $t$.  The coefficients that represent a constant hazard function can be derived as a deterministic function of the knots.  The full definition is given in the Appendix. 

The parameters $\eta$ and $p_i$ do not have exact interpretations --- the intention is to obtain a function that adapts to fit the data, rather than to learn the biological or clinical mechanism.  Three examples of M-spline hazard functions, on the same basis and scale, but with different coefficients, are illustrated in Figure~\ref{fig:mspline}.  

\subsection{Modelling data with an M-spline model}

The knots should be chosen to allow the hazard function to be determined from the data, and to allow the shape of the function to change outside the data, if this is plausible and extrapolation is needed.   This section explains the default approach in the \texttt{survextrap} package, but these defaults can be modified by placing knots anywhere if needed. 

As in~\citet{royston:parmar}, knots are placed by default at quantiles of the uncensored individual-level survival times.   If there are also external data, additional knots should be defined by the user if required to cover the times of these data.  The appropriate number and location of additional knots depends on how many external datasets there are and what times they cover --- noting that hazard changes \emph{within} an interval $(u_j,v_j)$ cannot be identified from an aggregate count of survivors over this interval.  Then to allow for hazard changes outside the period covered by all data, the highest knot should be placed at a point beyond which either we assume the hazard will not change, or any hazard changes are unimportant.

The appropriate level of flexibility for the hazard function is determined automatically from the data, by using a principle of \emph{penalisation}~\citep{wood2017generalized}, or ``shrinkage'' towards a constant hazard.  Firstly, the number of knots spanning the individual data is fixed to be large enough to accommodate all plausible shapes ($n=10$ basis functions is the current package default).  A hierarchical prior is then placed on the coefficients $p_i$, to represent a belief that the true model \emph{may} be this flexible, but is most likely to be less flexible.  Then, when the model is fitted to data, the resulting posterior represents the optimal level of flexibility needed to describe the data.   This is intended to protect against the risk of over-fitting.

Specifically, the prior for the $p_i$ is a multinomial logistic distribution: $\log(p_i / p_1) = \gamma_i$, with $\gamma_1=0$ and $\gamma_i \sim Logistic(\mu_i, \sigma)$ for $i = 2, \ldots, n$.   The prior mean $\boldsymbol{\mu} = (\mu_2,\ldots,\mu_n)$ is defined so that the corresponding $p_i$ represent a constant hazard $h(t)$.  The prior variability $\sigma$ controls the smoothness of the fitted hazard curve, and is estimated from the data.   If $\sigma=0$ then we are certain that the hazard is constant, and values of $\sigma$ around 1 favour ``wiggly'' curves where the hazard is driven mainly by the data.  See the \texttt{survextrap} package vignettes for some examples.

\subsection{Modelling explanatory variables}
\label{sec:covs}
To extend this model to allow the hazard $h(t \vert \x) = \eta \sum_i p_i b_i(t)$ to depend on explanatory variables $\x$, a proportional hazards model can be used, where the scale parameter $\eta$ is redefined as $\eta(\x) = \eta_0 \exp(\boldsymbol{\beta}^T \x)$.

A novel, flexible non-proportional hazards model is also defined, by also modelling the spline coefficients $p_i$ as functions of $\x$ using multinomial logistic regression: 
\begin{eqnarray*}
  \label{eq:nonprop}
  \log(p_i(\x) / p_1(\x)) & = & \gamma_i(\x) \\
  \gamma_i(\x) & \sim & Logistic(\mu_i + \boldsymbol{\delta}^{T}_i \x, \sigma) \quad (i=2,...,n), \quad \boldsymbol{\delta}_1 = 0 
\end{eqnarray*}
The $s$th element of the vector $\boldsymbol{\delta}_i$ describes the departure from proportional hazards for the $s\mbox{th}$ covariate in the region of time associated with the $i$th spline basis term.   With all $\boldsymbol{\delta}_i = 0$, this is the proportional hazards model. For each covariate $s$, a hierarchical prior is used for the non-proportionality effects, $\delta_{is} \sim Normal(0, \tau_s)$, which ``partially pools" or smooths the effects from different regions of time $i$.

Hence, in the non-proportional hazards model, the ratio between the hazards $h(t \vert \x)$ at different covariate values $\x$ is a function of time $t$.  Since the $\boldsymbol{\delta}_i$ may be arbirtarily different in each region of time $i$, the hazard ratio can be an arbitrarily-flexible function of time.  The shape of this time-dependence is estimated from the data, while the hierarchical prior protects against over-fitting.

\subsection{Special mechanisms}
\label{sec:mechanisms}

The idea behind the package is for inferences to be driven by transparently-stated data and judgements. As described so far, the flexible model used to achieve this is treated as a ``black box'', that is, its specific form is not intended to have a biological or clinical interpretation.  However, sometimes plausible mechanisms can be specified via parametric model structures,  to improve estimates of long-term survival.  The package currently supports the following three mechanisms, that are sometimes assumed for survival extrapolation. 

\paragraph{Additive hazards / relative survival}
Here the overall hazard is assumed to be the sum of two hazards for different causes of death, as $h(t) = h_b(t) + h_c(t)$, where
\begin{itemize} 
\item $h_b(t)$ is the ``background'' hazard, assumed to be a known, piecewise-constant function of time.  This is often confidently known from national statistics on general mortality. 
\item $h_c(t)$ is the ``cause-specific'' or ``excess'' hazard for the disease of interest, which is modelled with the M-spline parametric model.  
\end{itemize}
The individual-level data are assumed to be described by $h(t)$, and fitting the model to these data involves estimating the parameters governing $h_c(t)$. The corresponding survivor functions are multiplicative: $S(t)=S_b(t)S_c(t)$, hence the alternative term ``relative survival'' model.  This is a variant of the model used by \citet{nelson2007flexible}, but Bayesian and with a different kind of spline.  

\paragraph{Mixture cure model}
In a mixture cure model, data are assumed to arise from a survival function $S(t \vert p, \boldsymbol{\theta}) = p + (1-p)S_0(t\vert\boldsymbol{\theta})$, where the unknown parameter $p$ is sometimes termed the ``cure probability", and $S_0(t \vert \boldsymbol{\theta})$ is a parametric model with parameters $\boldsymbol{\theta}$, termed the ``uncured survival".  Here, $S_0$ is the M-spline model described above.  For very large $t$, the survival converges to a constant $p$, the probability of never experiencing the event.  This is often interpreted as a mixture of two populations, where a person has zero hazard at all times $t$ with probability $p$, or a hazard $h_0(t)$ at all times with probability $1-p$.   However, contrary to some descriptions of this model, this is not a necessary assumption, because the same survival function arises if everyone is subject to the same hazard that decreases asymptotically to zero over time, $h(t) = f_0(t) / \left(p/(1-p) + S_0(t) \right)$, where $f_0(t)$ is the probability density function of the ``uncured'' model.  These two interpretations are indistinguishable in practice, since in the mixture interpretation, we cannot determine which of the two populations censored observations belong to.

A mixture cure model can either be used for the overall hazard, or the cause-specific hazard $h_c$ in an additive hazards model.  Using a cure model for $h_c$ would be appropriate if we can assume that the cause-specific hazard decreases to a negligible amount over time, so that everyone with the disease of interest is essentially ``cured'', and their hazard becomes dominated by background causes.

\paragraph{Waning treatment effects}
Health technology assessments are often primarily informed by trials of a novel treatment against a control.  Beyond the trial horizon, information about the relative effect of the new treatment will be weak.  A naive extrapolation from the model would assume that the estimated short-term effect will continue in the long term (e.g. as a constant hazard ratio).  This is often contrasted with more conservative scenarios where the treatment effect wanes over time, so that after some point, treated and control patients have the same hazard. 

Treatment effect waning can be achieved by firstly fitting a parametric model $h(t\vert x)$, including the effect of a treatment $x$, to short-term data.  This does not necessarily need to be a proportional hazards model.  Then, the predicted hazard for the control group is taken from the fitted model, $h(t \vert x=0)$.  The predicted hazard for the treated group is obtained as $h(t \vert x=1) = h(t \vert  x=0) hr(t)$, where the time-dependent hazard ratio $hr(t)$ is defined as follows:
\begin{itemize}
\item For $t \leq t_{min}$, $hr(t)$ is taken from the fitted model. $t_{min}$ might be the end of the trial follow-up period, or a later point up to which the effect from the trial can be confidently extrapolated.
\item For $t \geq t_{max}$, $hr(t)=1$.
\item For $t_{min} \leq t \leq  t_{max}$, $\log(hr(t))$ is assumed to diminish linearly between $\log(hr(t_{min}))$ at $t_{min}$, and zero at $t_{max}$.  
\end{itemize}

\subsection{Bayesian inference}
\label{sec:bayes}
The models define a hazard function $h(t \vert \btheta,\x)$, from which the cumulative hazard function and survivor function $S(t \vert \btheta, \x)$ can be derived, to construct the likelihood function for the individual-level data.  This hazard function is also assumed to govern the external datasets $j$, with any differences between them explained by different explanatory variables $\x = \x_j$ and the time period covered.  The likelihood for each external dataset $j$ is built by assuming that $r_j$ is generated from a Binomial distribution with denominator $n_j$ and probability $p_j= S(u_j \vert \btheta, \x_j) / S(t_j  \vert \btheta, \x_j)$, that is, the probability of survival to time $u_j$ conditionally on being alive at $t_j$.  Samples from the posterior distribution of $\btheta$ (which may comprise e.g. $\eta$, $p_1,\ldots,p_n$, $\sigma$, $\boldsymbol{\beta}$, $\boldsymbol{\delta}$, $\tau_s$, depending on the model choice) are then obtained by Markov Chain Monte Carlo methods, specifically Hamiltonian Monte Carlo as implemented in Stan~\citep{stan}.

In the package, all priors for the parameters comprising $\btheta$ can be defined by the user.  If any priors are not specified, then the following defaults are used.  These are ``weakly informative'', that is, containing some stabilising information but largely letting the data drive the inferences, following principles described by~\citet{gelman2020regression}.  As in \citet{brilleman2020bayesian}, the baseline log hazard $\log(\eta_0)$ (for covariate values of zero) is given a normal prior with mean zero and standard deviation 20.  For the log hazard ratios, Normal(0, 2.5) is used, and a Gamma(2,1) is used for the smoothness parameter $\sigma$.

\paragraph{Prior calibration} Procedures are also provided for simulating from the joint prior distributions for the parameters in $\btheta$, to ensure that they imply plausible beliefs about easily-understandable quantities.  For example, $\sigma$ governs how much the hazard is expected to vary through time --- a constant hazard has $\sigma=0$, but other values of $\sigma$ are hard to interpret.  However, we can draw a simulated value from any given prior for $\sigma$, jointly with the the distributions for the $p_i$ and $\eta$, and deduce the implied hazard curve $h(t)$.  The hazard variability in this curve could be described by the ratio $\rho$ between (say) the 90\% percentile and 10\% percentile of $h(t)$ over a fine, equally-spaced grid of times $t$ from zero to the highest knot.  By repeatedly simulating hazard curves, we can draw from the prior distribution of $\rho$.  We can then calibrate the prior for $\sigma$ to achieve a prior on $\rho$ that expresses beliefs of the form ``the highest values of the hazard are unlikely to be 10 times the lowest values''.

\paragraph{Statistical model comparison} The goodness-of-fit of different models to the observed data can be compared using leave-one-out cross-validation, via the method and R package of~\citet{loocv,loor}.  For each observation $i$ (individual event or censoring time in the individual data, or individual event indicator in the external data), this method estimates $elpd_i$, the expected log predictive density, a measure of the accuracy with which a model would predict the $i$th observation if it were left out when fitting the model.  The sum LOOIC $= -2\sum_ielpd_i$ is then used as an ``information criterion'' to compare the fit of models.  LOOIC is similar in principle to DIC~\citep{spiegelhalter2002bayesian}, but with a direct interpretation in terms of predictive ability.

\section{Implementation of the software}
\label{sec:impl}

An R package called \texttt{survextrap} implements the method.  It is available from \url{https://chjackson.github.io/survextrap/}.  It uses the \texttt{rstan} interface to the Stan software~\citep{stan,rstan} to perform Hamiltonian Monte Carlo sampling from the posterior distribution of the Bayesian model.   Models can be fitted with a single R command, using a similar mechanism to the \texttt{rstanarm} package~\citep{brilleman2020bayesian}.  Posterior summaries (e.g. estimates and credible intervals) for a range of interesting outputs (e.g. survival probabilities, hazards, mean survival times) can then be extracted using single commands.  Outputs from all these functions obey ``tidy data'' principles~\citep{hadley2014tidy}, to facilitate further processing, in particular, plotting with the \texttt{ggplot2} R package.

An example of basic use of the package is given in the Appendix of this paper.  The website \url{https://chjackson.github.io/survextrap/} gives thorough documentation for all the package's functions, including a series of articles describing specific features in more detail, and the code and data to reproduce the analysis in Section~\ref{sec:demo}.

\section{Demonstration: cetuximab for head and neck cancer}
\label{sec:demo}

This section demonstrates a range of models that can be built with \texttt{survextrap} in a typical application to a health technology evaluation, each using different kinds of information and model assumptions to enable extrapolation.  The full R code to reproduce each model fit, calculation and plot is supplied and explained in a supplementary article.  The paper will focus on discussing different analysis choices and their consequences.  

We study the data that were previously analysed by~\citet{guyot2017extrapolation}, describing the survival of people with head and neck cancer.  Data are obtained from 5 years of follow-up of a trial \citep{bonner2006radiotherapy} of cetuximab and radiotherapy, compared to a control group who only received radiotherapy.  Individual-level data were imputed from the published Kaplan-Meier estimates, using the method of~\citet{guyot2012enhanced}.  As well as the trial data, there are two external data sources~\citep[full details are given in][]{guyot2017extrapolation}:
\begin{enumerate}
\item[(a)] a cancer registry (SEER: the Surveillance, Epidemiology and End Results Program), representing people with the same distribution of age, sex, cancer site and calendar period of diagnosis date as the trial data. This gave counts of survivors $r_j$ over the following year, out of $n_j$ alive at $j$ years after diagnosis, for $j=5$ to 25, giving estimates of annual survival probabilities $p_j$.
\item[(b)] survival data for the general population of the United States, matched by age, sex and date. 
\end{enumerate}
We examine either the survival for the control group alone, or the increase in survival provided by cetuximab.  Specifically we calculate the restricted mean survival time (RMST), or the difference in restricted mean survival times, over time horizons varying from 5 years up to 40 years.  We discuss how longer-term data and judgements are required to obtain confident estimates of longer-term survival. 

\subsection{Prior information}
To improve transparency, and stabilise computation, we specify priors explicitly, rather than relying on the package's (fairly vague) defaults.
\begin{enumerate}
\item Hazard scale parameter $\eta$.  Since patients in the trial have a median age of 57 (range 34 to 83), the prior for $\eta$ is calibrated to imply a prior mean survival of 25 years after diagnosis but with a variance chosen to give a wide 95\% credible interval of about 5 to 100 years. Note that this interval describes \emph{uncertainty} about knowledge of the \emph{mean} in the control group, not \emph{variability between individuals} in this group.

\item Smoothing parameter $\sigma$.  This is chosen by simulation (as described in Section~\ref{sec:bayes}) so that the highest hazard values for the control group (90\% quantile) are expected to be $\rho=2$ times the lowest values (10\% quantile), but with a credible interval for $\rho$ of between 1 and 16.
\end{enumerate}
The individual-level data are spanned by $n=6$ spline basis terms, chosen according to the quantiles of observed event times.  Beyond the trial data, additional knots are used depending on what external data are included, and what time horizon we want to estimate survival for.

\subsection{Trial data alone: extrapolating a single arm}
\label{sec:casestudy:singlearm}

\begin{figure}
  \centering
  \scalebox{0.58}{\includegraphics{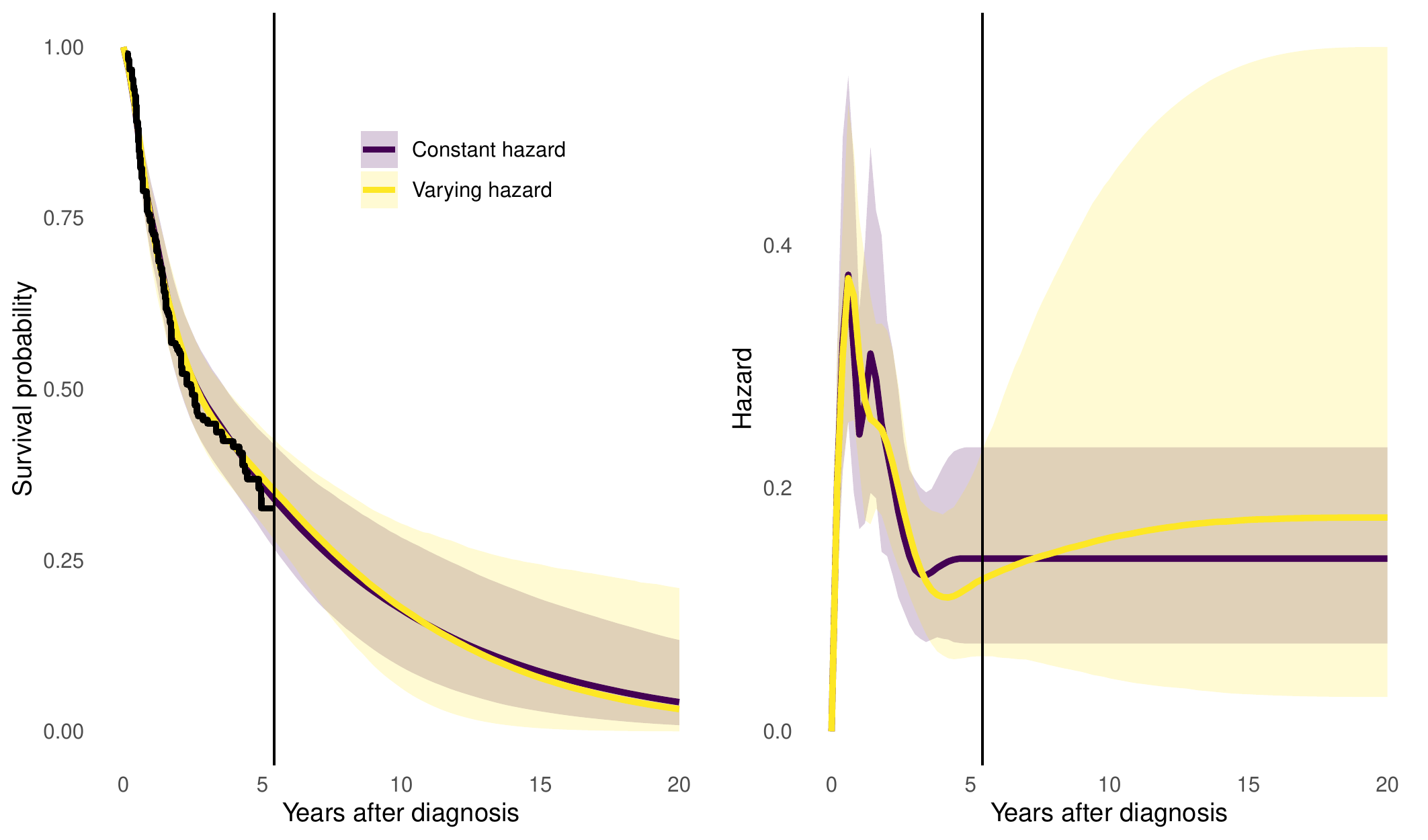}}
  \caption{Survival and hazard estimates (posterior median and 95\%
    credible intervals) for the trial control group, from flexible
    Bayesian models where the hazard after the last event time in the
    trial (vertical line) is either arbitrarily varying (1a) or constant (1b).
    Kaplan-Meier estimate from the trial data also shown on the left
    as a black jagged line.}
  \label{fig:control}
\end{figure}

Firstly we study just the data from the control treatment group.   We contrast two models that describe the trial data in the same way, but differ in how extrapolation is done, labelled as:
\begin{enumerate}
\item[(1a)] the highest knot is set to 20 years,
\item[(1b)] the highest knot is set to the final event time in the data (5 years in this case).  This is the package default if an upper knot is not specified.
\end{enumerate}
The posterior distributions of the survival and hazard curves up to 20 years (Figure~\ref{fig:control}) show how extrapolation relies on both data and assumptions. Here there are no data describing 5 to 20 years.  In (1b) we made the strong assumption that hazards will remain constant after the trial horizon of 5 years.  In (1a) we assumed that the hazard will change smoothly after the trial, but using a spline model that allows any size and direction of change, not determined by the fit to the short-term data.  Therefore there is a lot of uncertainty around the extrapolated hazard function in (1a), but the extrapolation under (1b) is more confident.   The exact extent of uncertainty in (1a) will be sensitive to where knots are placed, though a rough uncertainty quantification may be sufficient to highlight the need for further information beyond that included in the trial.

5 and 20-year RMST estimates are shown in Table~\ref{tab:rmst}. Credible intervals for 20-year estimates are wide when we do not constrain the extrapolated hazard.  The 5 year RMSTs from model (1a) did not change by more than 0.1 years when the model was made more or less flexible (through the number of basis terms varying from 5 to 12).

\subsection{Trial data alone: treatment comparisons}
\label{sec:casestudy:trt}

Now we consider a comparison between treatment groups based on the trial data alone, using three alternative models, labelled as:
\begin{enumerate}
\item[(2a)] a proportional hazards model,
\item[(2b)] a parametric non-proportional hazards model (Section~\ref{sec:covs}),
\item[(2c)] both treatment arms modelled separately.
\end{enumerate}

\begin{figure}
  \centering
  \scalebox{0.58}{\includegraphics{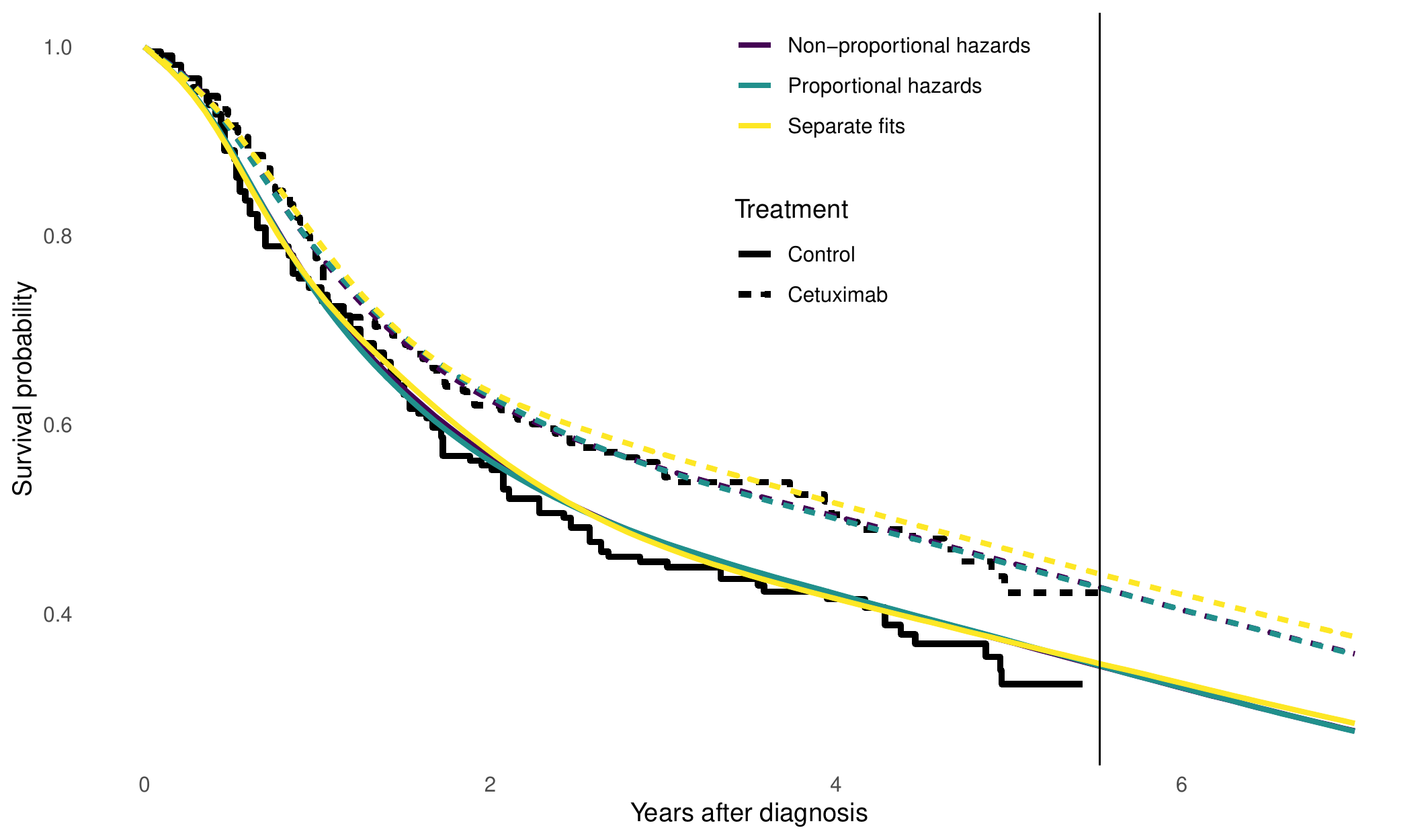}}
  \caption{Estimates of survival as posterior medians from three
    different flexible Bayesian models for both treatment groups.
    Kaplan-Meier estimates from the trial data are shown as jagged
    lines, and spline knot locations as vertical lines.}
  \label{fig:trt}
\end{figure}

These models fit similarly well to the trial data, judging from the fitted survival curves (Figure~\ref{fig:trt}), and their similar LOOIC cross-validation statistics (Table~\ref{tab:trt}).  The difference between them is more apparent when extrapolating.   The upper boundary knot is set to 20 years, so that we allow the hazard to change after 5 years, even though there is no data then.  Over five years (Table~\ref{tab:trt}) the survival and incremental survival between treatment groups is similar between the three models, but over 20 years the uncertainty about these quantities is greater.   The credible intervals are narrowest under the proportional hazards model, and widest when modelling arms independently.   The non-proportional hazards model makes more efficient use of the data than modelling arms separately, though the proportional hazards model is adequate (judging by LOOIC).

\begin{table}[htbp]
  \begin{tabular}{lllp{0.3in}p{1in}}
\hline
Model & Observed data  & Extrapolation assumptions & Time \newline horizon & Restricted mean \newline survival\\
\hline
(1a)  &  Trial  &  No extrapolation  &  5  &  2.88 (2.63, 3.15)\\
\hline
(1a)  &  Trial  &  Uncertain hazard  &  20  &  5.11 (3.81, 7.13)\\
(1b)  &  Trial  &  Constant hazard  &  20  &  5.14 (4, 6.73)\\
(1c)  &  Trial,registry  &  No extrapolation  &  20  &  5.76 (5.04, 6.57)\\
\hline
(1d)  &  Trial,registry  &  Uncertain hazard  &  40  &  6.18 (5.36, 7.17)\\
(1e)  &  Trial,registry,population  &  Uncertain excess hazard  &  40  &  6.21 (5.36, 7.12)\\
(1f)  &  Trial,registry,population  &  Mixture cure  &  40  &  6.28 (5.4, 7.23)\\
(1g)  &  Trial,registry,population  &  Elicited survival  &  40  &  6.26 (5.36, 7.22)\\
\hline
  \end{tabular}
  \caption{Comparison of estimates of restricted mean survival time in years (posterior median and 95\% credible intervals) over different models and time horizons, for head and neck cancer patients in the control group.  The models differ by the different sources of observed data included, and the different assumptions used for extrapolation outside the time horizon of the observed data.}
  \label{tab:rmst}
\end{table}

\begin{table}[h]
  \begin{tabular}{lp{1.3in}p{1.3in}l}
\hline
Model  &  Restricted mean survival \newline (control)  &  Increase in mean survival \newline (cetuximab - control)  &  LOOIC\\
\hline
\multicolumn{4}{l}{Trial data alone: prediction horizon 5 years}\\
(2a) Proportional hazards  &  2.87 (2.63,3.12)  &  0.31 (-0.03,0.66)  &  1156\\
(2b) Non-proportional hazards  &  2.88 (2.62,3.12)  &  0.3 (-0.07,0.67)  &  1158\\
(2c) Separate arms  &  2.88 (2.63,3.15)  &  0.36 (0,0.72)  &  1160\\
\hline
\multicolumn{4}{l}{Trial data alone: prediction horizon 20 years}\\
(2a) Proportional hazards  &  4.92 (3.71,6.67)  &  1.14 (-0.09,2.68)  &  \\
(2b) Non-proportional hazards  &  4.93 (3.7,6.99)  &  1.08 (-0.64,2.86)  &  \\
(2c) Separate arms  &  5.11 (3.81,7.13)  &  1.33 (-1.23,4.1)  &  \\
\hline
\multicolumn{4}{l}{Trial and registry data: prediction horizon 20 years, proportional hazards models}\\
(2d) No waning  &  5.81 (5.05, 6.55)  &  1.21 (-0.38, 2.9)  &  \\
\hline
\multicolumn{4}{l}{Trial, registry and population data: prediction horizon 20 years, proportional hazards models}\\
(2e) No waning  &  5.87 (5.07, 6.66)  &  1.11 (-0.52, 2.72)  &  \\
(2e) 5 to 20 years  &  5.87 (5.07, 6.66)  &  1.06 (-0.5, 2.56)  &  \\
(2e) 5 to 6 years  &  5.87 (5.07, 6.66)  &  0.84 (-0.4, 2.01)  &  \\
\hline
  \end{tabular}
  \caption{Comparison of models fitted to both treatment and control trial data.  Estimates of restricted mean and increase in restricted mean survival in years over 5 or 20 year horizons (posterior median and 95\% credible intervals), and LOOIC model comparison statistic (lower indicates better predictive ability)}
  \label{tab:trt}
\end{table}
All the models so far have ignored the substantive information that exists beyond the trial data: the registry and population data to inform mortality for these patients, and information about the mechanism of the treatment effect.   

\subsection{External data from the patients of interest}
\label{sec:casestudy:external}
\begin{figure}
  \centering
  \scalebox{0.58}{\includegraphics{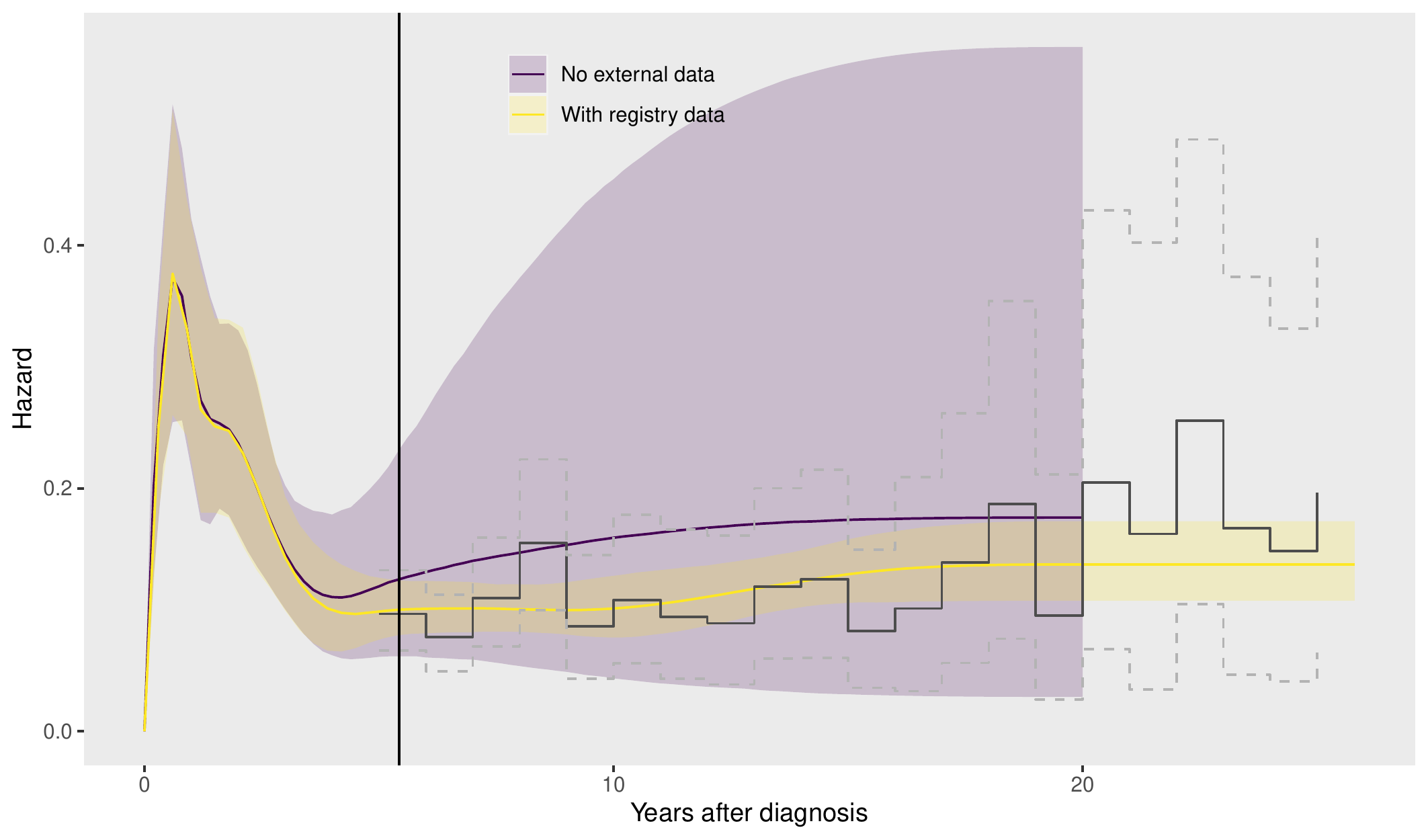}}
  \caption{Hazard estimates (posterior median and 95\% credible intervals) from models for trial data with (1c) and without (1a) external data from the cancer registry (coloured bands), and hazard estimates from the registry data alone (jagged lines)}
  \label{fig:registry}
\end{figure}

We now examine how to incorporate external data in \texttt{survextrap} models.  Annual hazard (mortality rate) estimates from the SEER registry data, calculated as $-\log(r_j/n_j)$, are illustrated in Figure~\ref{fig:registry}, with corresponding interval estimates (calculated from quantiles of the $Beta(r_j,n_j-r_j)$).   These are included in a joint model with the trial data (labelled (1c) in Table~\ref{tab:rmst}), with knots added at 10, 15 and 20 years, and the patients in the registry are assumed to have the same survival as the control group of the trial.  The posterior distribution of the hazard from this model is also illustrated in Figure~\ref{fig:registry}, along with estimates from the equivalent model (from Figure~\ref{fig:control}) that excludes the registry data.   The registry data makes the extrapolated hazard and RMST much more confident.   The model allows the hazard to vary flexibly up to 20 years, and those variations can be estimated from the registry data.  Different knot placements did not substantially affect estimates of survival over 20 years or improve the fit to the external data as measured by LOOIC (see the supplementary material for more details).

Note that in the \texttt{survextrap} package, the external data does not have to describe a population identical to that described by a particular subgroup of the individual data: the differences between data sources could instead be explained by covariates included in the model, using either proportional or non-proportional hazards.

\subsection{Population data informing background mortality} 

Another way of including external data is through additive hazards, as described in Section~\ref{sec:mechanisms}.  Here this allows the data on survival of the general population to be included.  These are assumed to follow the background hazard $h_b(t)$, which is assumed known.  The trial data follow the overall hazard $h(t)$, and the excess hazard $h_c(t)$ for head and neck cancer patients is assumed to follow the flexible M-spline model and estimated.   This model constrains the survival of head and neck cancer patients to be no better than the survival of the general population. 

The population data are added to the model that includes the registry data.  We compare hazard extrapolations up to 40 years, placing further knots at 30 and 40 years (in addition to those spanning the trial and registry data), either without or with the population data (Figure~\ref{fig:registry}), labelled (1d) and (1e) in  Table~\ref{tab:rmst}.  The population data do not affect the hazard estimates up to 20 years, but the extrapolations over 40 years are very uncertain unless the population data are included.  Including the population data allows the reasonable constraint that hazards will not go below those of the general population.  The exact excess risk for head and neck cancer patients is still uncertain, however, since we do not have data beyond 25 years to inform it.

\begin{figure}
  \centering
  \scalebox{0.58}{\includegraphics{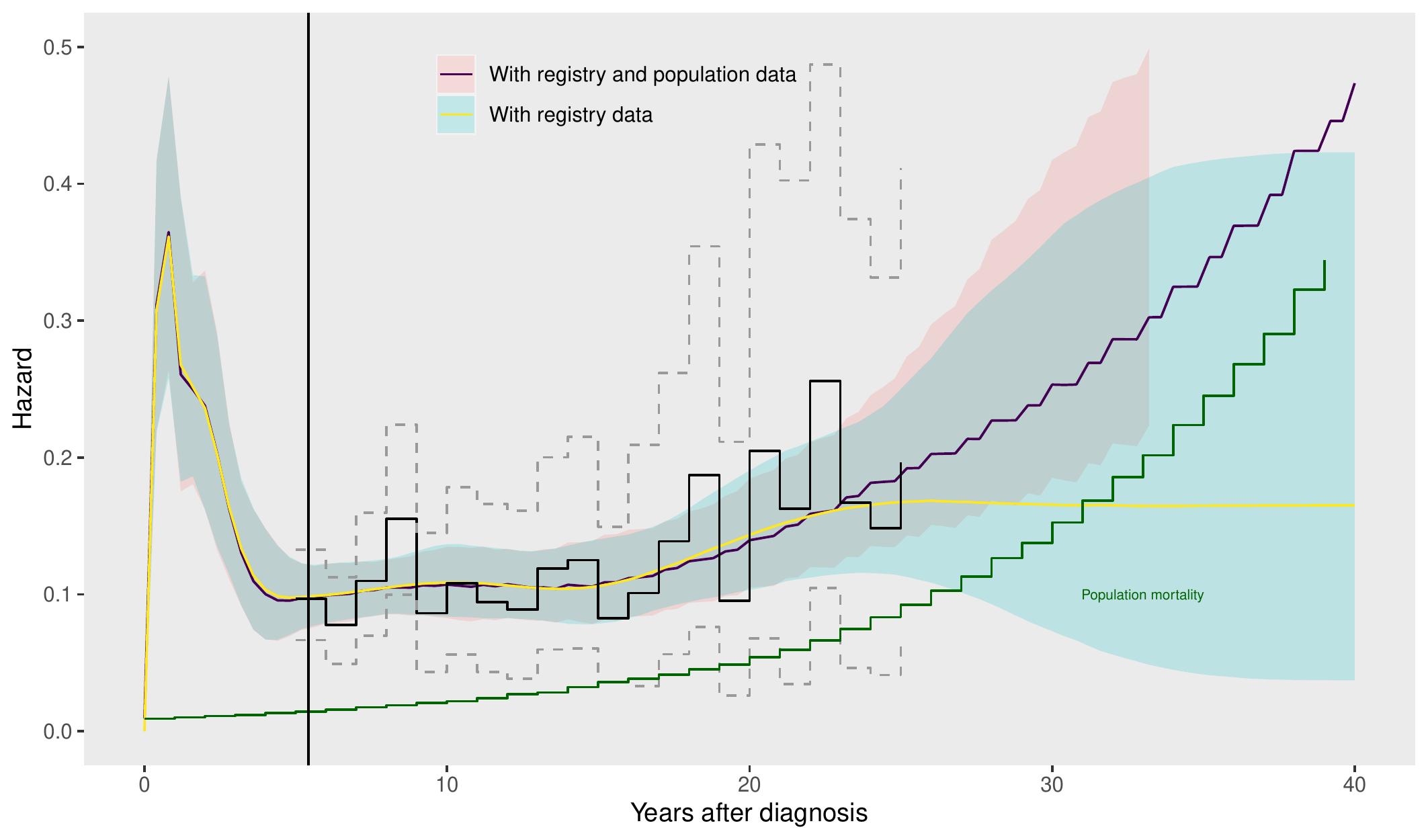}}
  \caption{Hazard estimates (posterior median and 95\% credible intervals) from models from the trial data and cancer registry data, compared without (1d) and with (1e) additional data from the general population. }
  \label{fig:pop}
\end{figure}

\subsection{Mixture cure model}

\begin{figure}
  \centering
  \scalebox{0.58}{\includegraphics{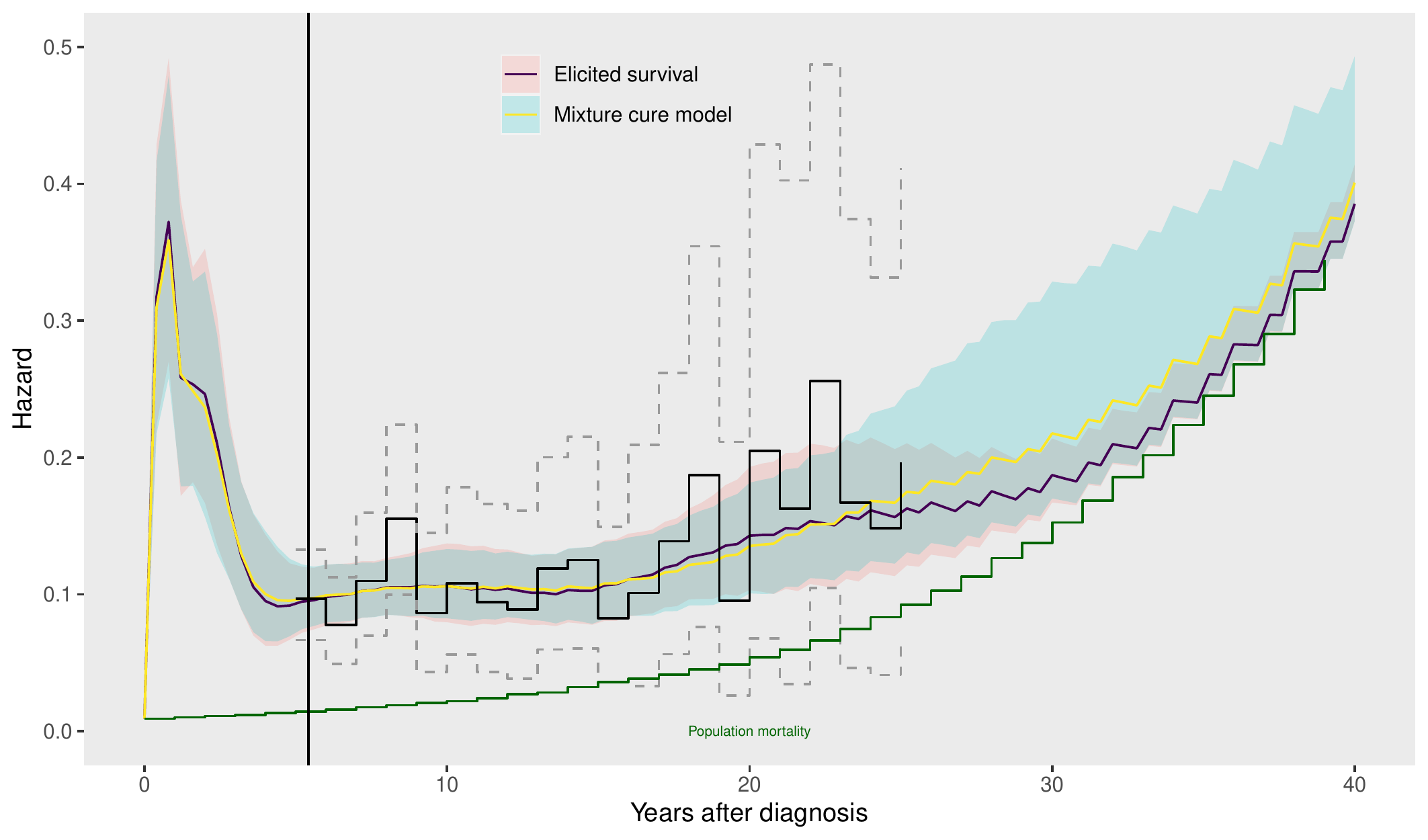}}
  \caption{Hazard estimates (posterior median and 95\% credible intervals) from models from the trial, cancer registry and general population data under two alternative models ((1f) mixture cure, and (1g) elicited survival) that represent a decreasing excess hazard for people diagnosed with head and neck cancer.}
  \label{fig:cure}
\end{figure}

We could improve the precision of the estimates of the excess hazard for head and neck cancer patients by including judgements, for example, that the excess hazard will diminish to zero as people age.  While there is no evidence of this from the registry data in this example, in some applications it might be plausible.  One way to represent this might be through a mixture cure model (Section~\ref{sec:mechanisms}) fitted to the trial, registry and population data combined.    Comparing the results from the mixture cure model (Figure~\ref{fig:cure}, and (1f) in Table~\ref{tab:rmst}) to the model for the same data with no cure assumption (Figure~\ref{fig:pop} and (1e)) shows how the assumption of cure has pulled the hazard extrapolations for 20-40 years closer to the estimates of the background hazard, though with wide credible intervals.   The exact shape of the extrapolation for the cure model is influenced by the parametric form for the mixture cure hazard function.  In practice, this should be checked for plausibility. 

\subsection{Elicitation of long-term survival probabilities}

A more flexible way to include longer-term judgements is by eliciting survival probabilities.   These can be converted to artificial datasets represeting counts of survivors, which can be included as ``external data'' in the model, using the idea described in Section~\ref{sec:model}. 

For example, we could state an assumption of ``cure'' in the form: ``by 40 years after diagnosis, we are confident that the patients of interest will have similar mortality to the background population''.  The annual survival probability in the matched general population dataset at this point is 0.72.  A $Beta(1000\times0.72, 1000\times(1-0.72)$ distribution, $Beta(724,276)$, has a 95\% credible interval of (0.70, 0.75).  This is equivalent to having observed $r_j=724$ survivors at the end of the year, from $n_j=1000$ people alive at the start.  The denominator $n_j$ could be controlled to give different amounts of prior uncertainty, e.g. $n_j=100$ would give a $Beta(72,100-72)$ which has a wider credible interval of $(0.63, 0.80)$. 

This artificial dataset is then concatenated with the SEER registry data, and supplied to the model in the same way as the registry data.   The predicted hazard from this model is shown in Figure~\ref{fig:cure}.  This reflects our assumption that the overall hazard approaches the background hazard at 40 years, but with some uncertainty.     This model makes less restrictive parametric assumptions than the mixture cure model --- since spline knots are placed at 20, 30 and 40 years, the hazard curve is allowed to change within a wide variety of smooth shapes.  The assumption that the cancer patients are ``cured'' by 40 years is provided through a directly-stated judgement about survival at 40 years, rather than through extrapolating a parametric equation estimated only from data up to 25 years.

Finally, note that the RMST estimates (Table~\ref{tab:rmst}) do not change much between the four different assumptions (1d)--(1g) used for estimating the hazard between 20 and 40 years, since the probability of survival beyond 20 years is low.

\subsection{Waning treatment effects}

We have now built in a model that includes all background information about general mortality of the patients in the trial, allowing us to extrapolate absolute survival of the control group as confidently as the data allow.    In Section~\ref{sec:casestudy:trt} we built models to estimate treatment effects from the trial data.    We now consider what judgements might be made about the treatment effect beyond the trial horizon, and how these can be modelled with  \texttt{survextrap}. 

As discussed by~\citet{guyot2017extrapolation}, the mechanism of cetuximab is to enhance the effect of radiotherapy.  The effects of both of these therapies is expected to be limited to the initial 5 or 6 years, where most of the mortality due to the cancer occurs. Therefore~\citet{guyot2017extrapolation} judged that the hazard ratio for the effect of adding cetuximab to radiotherapy is expected to diminish to one by around 6 years, though acknowledged some uncertainty around this.

Therefore the model which includes registry and population data but no cure is extended to include a proportional hazards model for treatment (the treatment effect mechanism that best fitted the trial data in Section~\ref{sec:casestudy:trt}).  The results from this model are labelled (2e) in Table~\ref{tab:trt}. The incremental restricted mean survival over 20 years is compared between three different assumptions about the treatment effect beyond the 5-year trial horizon:
\begin{enumerate}
\item[(a)] the log hazard ratio remains constant at the value estimated from the trial 
\item[(b)] the log hazard ratio wanes linearly from the trial value at 5 years to zero at 6 years
\item[(c)] the log hazard ratio wanes linearly from the trial value at 5 years to zero at 20 years
\end{enumerate}
Using all three data sources, with no waning, the incremental restricted mean survival over 20 years is estimated as 1.11 (-0.52, 2.72). This reduces to 1.06 (-0.5, 2.56) when waning is applied gradually from 6 to 20 years, and even further to 0.84 (-0.4, 2.01) when the effect is assumed to wane rapidly from 5 to 6 years.  Note also that omitting the population data from this model ((2d) in Table~\ref{tab:trt}) impacts the estimated treatment effect.

The assumptions made here are uncertain, and there are many ways in which this uncertainty could be described.  A simple deterministic sensitivity analysis is done here, which has an advantage of transparency to decision-makers.  An alternative approach would be to represent this uncertainty probabilistically (see, e.g~\citet{guyot2017extrapolation} for one approach), though formally specifying and eliciting distributions for a weakly-informed, time-varying quantity like this is challenging in general.

\section{Discussion}
\label{sec:discussion}
  
This paper has introduced a tool that makes principled methods for survival extrapolation straightforward.  It accommodates a wide range of data sources, that can be represented in a flexible statistical model.  The Bayesian approach allows uncertainty to be quantified, and the R package removes the need for specialised programming.  The model can represent uncertainty about how the hazard will change beyond the data, assuming only that the hazard function is smooth.

While the model is flexible, all models are based on assumptions.  The package tries to make these as transparent as possible.  In particular, prior distributions can easily be chosen to represent beliefs about interpretable quantities.  While the spline model relies on a choice of knots, the statistical fit of different choices to data can be compared.  Extrapolations outside data may be sensitive to modelling choices, but uncertainty is inevitable when data is weak.  If there is uncertainty, there is a tension between decision-making and recommending collection of further data.  The Bayesian approach represents uncertainty using probability distributions, which allows the use of ``value of information'' methods to estimate the expected benefits of further information (e.g. from a health economic perspective).  In principle, the posterior distribution from a \texttt{survextrap} model might be used to calculate the expected value of sample information for a new trial or further follow-up from an existing trial --- the implementation details have not been worked out, but see e.g.~\citet{vervaart2023general} for a potential starting point.

There are several ways in which the model used here might be extended.  \emph{Hierarchical} or random effects models are one potential direction, as in \texttt{rstanarm}~\citep{brilleman2020bayesian}.  These might be used to represent various kinds of heterogeneity in survival, e.g. between observed groups such as different hospitals~\citep{ieva2017multi}, or between latent classes of individuals~\citep{federico2022heterogeneity}.  Survival models with random effects can also be used for (network) meta-analysis~\citep{jansen2011network}.  Another common extension of survival models is to \emph{multi-state} models for times to multiple events.  See, e.g.~\citet{jackson2022comparison} for a comparison of flexible parametric frameworks for multi-state models, and~\citet{jansen2020multi} for network meta-analysis of survival data with multiple outcomes.  The ideas described in this paper would enable any of these previous methods to be strengthened by including background information from external data.

A final point to consider is that getting new statistical methods into routine practice involves several ``phases'' of research~\citep{heinze2022phases}.  This paper has described the theoretical basis for a novel method, shown its utility in a realistic application, and provided software to make it usable with the minimum of effort.   However, to improve confidence in it, more work using it in a wider range of applications, and perhaps simulation studies, would be helpful.  Education in statistical skills is also important.  The Bayesian spline models used here are more complex than basic parametric survival models, with many ingredients that may be unfamiliar, such as prior distributions, spline knots and Markov Chain Monte Carlo computation.  The online documentation includes lots of worked examples to explain the important concepts, and will be updated as a ``live'' resource in response to users' needs if the package becomes more widely used.

\paragraph{Acknowledgements}

I am grateful to Iain Timmins for suggesting the smoothness constraint at the spline boundary, and other helpful discussion.  Thanks also to Nicky Welton, Mike Sweeting, Dawn Lee, Ash Bullement, Nick Latimer, Ed Wilson, Gianluca Baio and Howard Thom for discussions and encouragement, and to Daniel Gallacher regarding the package name.    Funding was from the Medical Research Council, programme number MRC\_MC\_UU\_00002/11.  A Creative Commons licence will be applied to any accepted version of this manuscript.

\paragraph{Software availability and requirements}
\begin{itemize}
\item Project name: \texttt{survextrap}: an R package for survival extrapolation with a flexible parametric model and external data
\item Project home page: \url{https://chjackson.github.io/survextrap}
\item Operating system(s): Windows, MacOS and Linux
\item Programming language: R and Stan
\item Other requirements: R and various R packages, installed automatically
\item License: GNU GPL ($\geq 3$)
\end{itemize}

\paragraph{Availability of data and materials}
All data analysed during this manuscript are made available inside the \texttt{survextrap} package.  A detailed article explaining the case study in section~\ref{sec:demo}, with embedded R code to directly reproduce all results including graphs and tables, is available in the supplementary file \verb+cetuximab.html+, and in a ``live'' version at \url{https://chjackson.github.io/survextrap/articles/cetuximab.html} which will keep it up to date with any future enhancements or fixes to the software.


\section*{Appendix}

\subsection*{Rationale for using the M-spline model}
\label{app:mspline:rationale}

An advantage of generalised additive models such as splines, compared to other families of flexible models that have been used for survival analysis, e.g. those based on Gaussian processes~\citep{fernandez2016gaussian} or Dirichlet processes~\citep{deiorio:nph}, is that they express a flexible function in an easily-computable form, as a linear function of basis terms.  In the M-spline model, both the hazard and the cumulative hazard, required to compute the likelihood, are defined as linear functions of basis terms.

Many other kinds of spline or basis models are available, however. The model of \citet{royston:parmar}, which defines the log cumulative hazard as a natural cubic spline function of log time, is often used for survival analysis, but this model does not inherently impose the required constraint that the cumulative hazard is an increasing function of time.  To fit the model to data by maximum likelihood, the constraint is imposed while fitting, by rejecting proposed combinations of parameter values that are found to give decreasing cumulative hazards.  However, this approach is problematic for a Bayesian model, which requires a prior distribution specified in advance of estimation.  The prior should represent knowledge about plausible parameter values, and it is unclear how a prior that excludes invalid combinations of parameters would be defined in advance.

This problem would be avoided if the spline model were placed on the log \emph{hazard}~\citep[as in][]{crowther2014general}, which does not require a constraint to be positive or increasing.  However, expensive numerical integration would generally then be required to compute the cumulative hazard.  The advantage of the M-spline is that it can be placed directly on the hazard, as it is designed to represent a non-negative function, and it integrates analytically so that the cumulative hazard can be computed easily.

\subsection*{M-spline construction}
\label{app:mspline}

As defined in Section~\ref{sec:mspline}, we model a hazard function as $$h(t) = \eta \sum_{i=1}^n p_i b_i(t)$$ where $\eta>0$ and $p_i: i=1,\ldots,n, \sum_i p_i =1$ are parameters to be estimated, and the $b_i(t)$ are deterministic functions of time $t$, known as ``basis functions''.   The basis functions are defined using an extension of the standard ``M-spline'' basis. 

\paragraph{Standard M-spline basis}

The standard M-spline basis, as described by~\citet{ramsay1988monotone} after~\citet{curry1966polya}, is designed to build flexible functions $h(t)$ of a variable $t$, where $t$ is in a finite interval $[L,U]$.  To construct this, a grid of points  $t_1,\ldots,t_{n+k}$ on this interval is defined, where
\begin{itemize}
\item $t_1=\ldots=t_k$ is the ``lower boundary'' $L$,
\item $t_{k+1}$ up to $t_n$ comprise the ``internal knots'',
\item $t_{n+1}=\ldots=t_{n+k}$ is $U$, the ``upper boundary'' knot,
\item $k$ is the ``order'' of the basis.  This governs the degree $d$ of the polynomials that are used to define the basis.  $d = k-1$, so that if $k=2$, then the basis functions will be piecewise linear ($d=1$), and if $k=4,d=3$, the functions will be built from cubic polynomials. 
\end{itemize}
Given an order $k$, the $i$th basis term $b_i(t|k)$ for $i = 1, \ldots, n$ is defined recursively as follows.  Firstly the $b_i(t|k)$ are defined for $k=1$ as 
\[ b_i(t | k=1) = 1 / (t_{i+1} - t_i)
  \begin{array}{l}
    \mbox{if $t$ is between the knots }t_i, t_{i+1}\\
    \mbox{0 otherwise}
  \end{array}
\]
Then for each subsequent order $k$, $b_i(t | k)$ are defined in terms of the basis functions for the previous order $k-1$, as  
\[ b_i(t  | k) = \frac{k \left[ (t - t_i) b_i(t | k-1, t) + (t_{i+k} - t) b_{i+1}(t | k-1, t) \right]}{(k-1)(t_{i+k} - t_i)} \] 

For example, this defines $b_i(t|k=2)$ to be triangular functions on $(t_i,t_{i+2})$ with mode at $t_{i+1}$.  Whatever the degree, $b_i(t)$ are positive in $(t_i, t_{i+k})$, zero elsewhere, and integrate to 1 (see, e.g. Figure~\ref{fig:mspline}).

If $p_i = \frac{t_{i+k} - t_i}{k(U - L)}$ for $i=1,\ldots,n$, then $\sum_{i=1}^n p_i b_i(t)$ is constant with time, and $p_ib_i(t)$ becomes a basis function for a ``B-spline''~\citep{ramsay1988monotone}, which is used for modelling unrestricted functions.

\paragraph{Extension to model hazard functions}

Here this standard M-spline basis is extended to define a spline that can represent hazard functions $h(t)$ for time-to-event analysis.   The standard basis already satisfies the requirement that $h(t)\geq 0$, but we also require $h(t)$ to be defined for any time $t \geq 0$, rather than just a finite interval.  To achieve this: 
\begin{enumerate}
\item the lower boundary is fixed to $L=0$,
\item the value of $h(t)$ for all $t>U$ is taken to be constant, and defined by the value $h(U)$.
\end{enumerate}

The ``knots'' comprise the internal knots and the upper boundary $U$.  These are chosen by the user to give an appropriately-flexible function for their application.

\paragraph{Further extension for smoothness at the boundary}

We can also improve the function's smoothness around the upper boundary $U$ by restricting the derivative and second derivative at $U$ to both be zero.  Without loss of generality, suppose that the scale parameter $\eta = 1$, so that $h(t) = \sum_i p_i b_i(t)$.   To deduce the simplified basis, we first note that, for degree $d=3$, the $h(t)$ and its derivatives at the upper boundary $U$ take the simple form
\begin{eqnarray*}
h(U)  & = &  p_n b_n(U)\\
h'(U)  & = &  p_{n-1}b'_{n-1}(U) + p_n b_n'(U)\\
h''(U)  & = &  p_{n-2}b''_{n-2}(U) + p_{n-1}b''_{n-1}(U) + p_n b_n''(U)  
\end{eqnarray*}
since the other $b_i(t)$ and their derivatives are zero at $U$. 
Therefore if $h'(U)= h''(U)=0$, we can deduce that 
\begin{eqnarray*}
p_n  & = &  C/b_n(U)\\
p_{n-1}  & = &  (C/b_n(U)) (b_n'(U) / b'_{n-1}(U))\\
p_{n-2}  & = &  (C/b_n(U)) (b_n'(U) / b'_{n-1}(U)) b''_{n-1}(U) + (C/b_n(U)) b_n''(U)
\end{eqnarray*}
where $h(U)=C$.  This allows us to define a simpler basis with two fewer terms,
\[ h(t) = \sum_{i=1}^{n-2} p_i^* b_i^*(t), \]

by replacing the final three terms of the original basis $p_{n-2}b_{n-2}(t) + p_{n-1}b_{n-1}(t) + p_{n}b_{n}(t)$
with a single term $p_{n-2}^*b_{n-2}^*(t)$.   In this new term,
\begin{itemize}
\item the coefficient is $p_{n-2}^* = C$, easily interpreted as the hazard value at $U$, and
\item $b_{n-2}^*(t)$ is computed as a function of the original basis terms $b_n(t)$, $b_{n-1}(t)$, $b_{n-2}(t)$ and their values, derivatives and second derivatives at $U$.
\end{itemize}

\subsection*{Example of basic use of \texttt{survextrap}}

The dataset \texttt{colons} gives survival or censoring times of 191
patients from a trial of two chemotherapy regimes for colon cancer:
levamisole (\texttt{"Lev+"}) and levamisole combined with fluorouracil
(\texttt{"Lev+5FU"}), compared to an observation-only control group
(\texttt{"Obs"}). Suppose there are also external data, giving 50
survivors to 10 years out of 100 people alive at 5 years, and 40
survivors to 15 years out of 100 alive at 10 years.

\begin{Shaded}
\begin{Highlighting}[]
\NormalTok{extdat }\OtherTok{\textless{}{-}} \FunctionTok{data.frame}\NormalTok{(}\AttributeTok{start =} \FunctionTok{c}\NormalTok{(}\DecValTok{5}\NormalTok{, }\DecValTok{10}\NormalTok{), }\AttributeTok{stop =}  \FunctionTok{c}\NormalTok{(}\DecValTok{10}\NormalTok{, }\DecValTok{15}\NormalTok{),}
                     \AttributeTok{n =} \FunctionTok{c}\NormalTok{(}\DecValTok{100}\NormalTok{, }\DecValTok{100}\NormalTok{), }\AttributeTok{r =} \FunctionTok{c}\NormalTok{(}\DecValTok{50}\NormalTok{, }\DecValTok{40}\NormalTok{), }\AttributeTok{rx =} \StringTok{"Obs"}\NormalTok{)}
\end{Highlighting}
\end{Shaded}

\noindent We fit a proportional hazards model to the individual and
external data jointly. The prior for the log hazard ratio is changed
from its default, and assumed to apply to both treatment groups.

\begin{Shaded}
\begin{Highlighting}[]
\FunctionTok{library}\NormalTok{(survextrap)}
\NormalTok{npc\_mod }\OtherTok{\textless{}{-}} \FunctionTok{survextrap}\NormalTok{(}\FunctionTok{Surv}\NormalTok{(years, status) }\SpecialCharTok{\textasciitilde{}}\NormalTok{ rx, }\AttributeTok{data=}\NormalTok{colons,}
                      \AttributeTok{prior\_loghr =} \FunctionTok{p\_normal}\NormalTok{(}\DecValTok{0}\NormalTok{, }\FloatTok{1.5}\NormalTok{),}
                      \AttributeTok{external =}\NormalTok{ extdat)}
\end{Highlighting}
\end{Shaded}

Posterior summaries of the hazard ratios for each chemotherapy versus
control, restricted mean survival at 5 years, and survival
probabilities at 5 and 10 years, are produced. These are the median
and 95\% credible intervals by default, but any summary can be
produced.  The outputs are all data frames, with one row per quantity
of interest, and columns that define the quantity and give different
posterior summaries for it.  This is intended to obey ``tidy data''
principles, hence to facilitate further data processing and plotting,
e.g. with the \texttt{ggplot2} package.  100 iterations
\texttt{niter} from the posterior are used for a quicker but rougher
summary --- this should be increased if more precision is needed.

\begin{Shaded}
\begin{Highlighting}[]
\FunctionTok{summary}\NormalTok{(npc\_mod) }\SpecialCharTok{\%\textgreater{}\%} 
    \FunctionTok{filter}\NormalTok{(variable}\SpecialCharTok{==}\StringTok{"hr"}\NormalTok{) }\SpecialCharTok{\%\textgreater{}\%} 
    \FunctionTok{select}\NormalTok{(term, median, lower, upper)}
\end{Highlighting}
\end{Shaded}

\begin{verbatim}
## # A tibble: 2 x 4
##   term      median lower upper
##   <chr>      <dbl> <dbl> <dbl>
## 1 rxLev      0.719 0.427 1.14 
## 2 rxLev+5FU  0.542 0.319 0.926
\end{verbatim}

\begin{Shaded}
\begin{Highlighting}[]
\FunctionTok{rmst}\NormalTok{(npc\_mod, }\AttributeTok{t=}\DecValTok{3}\NormalTok{, }\AttributeTok{niter=}\DecValTok{100}\NormalTok{)}
\end{Highlighting}
\end{Shaded}

\begin{verbatim}
##        rx variable t median 2.5% 97.5%
## 1     Obs     rmst 3   2.02 1.82  2.19
## 2     Lev     rmst 3   2.27 2.00  2.47
## 3 Lev+5FU     rmst 3   2.42 2.18  2.61
\end{verbatim}

\begin{Shaded}
\begin{Highlighting}[]
\FunctionTok{survival}\NormalTok{(npc\_mod, }\AttributeTok{t=}\FunctionTok{c}\NormalTok{(}\DecValTok{5}\NormalTok{,}\DecValTok{10}\NormalTok{), }\AttributeTok{niter=}\DecValTok{100}\NormalTok{)}
\end{Highlighting}
\end{Shaded}

\begin{verbatim}
##          rx  t median lower upper
## 1       Obs  5  0.366 0.263 0.448
## 1.1     Obs 10  0.165 0.103 0.211
## 2       Lev  5  0.480 0.347 0.616
## 2.1     Lev 10  0.276 0.154 0.422
## 3   Lev+5FU  5  0.592 0.434 0.700
## 3.1 Lev+5FU 10  0.391 0.218 0.541
\end{verbatim}

The default plot method gives the estimated survival and hazard, and
optionally also uncertainty intervals around these.  

\begin{Shaded}
\begin{Highlighting}[]
\FunctionTok{plot}\NormalTok{(npc\_mod, }\AttributeTok{xlab=}\StringTok{"Years"}\NormalTok{, }\AttributeTok{niter=}\DecValTok{100}\NormalTok{)}
\end{Highlighting}
\end{Shaded}

\begin{center}\includegraphics{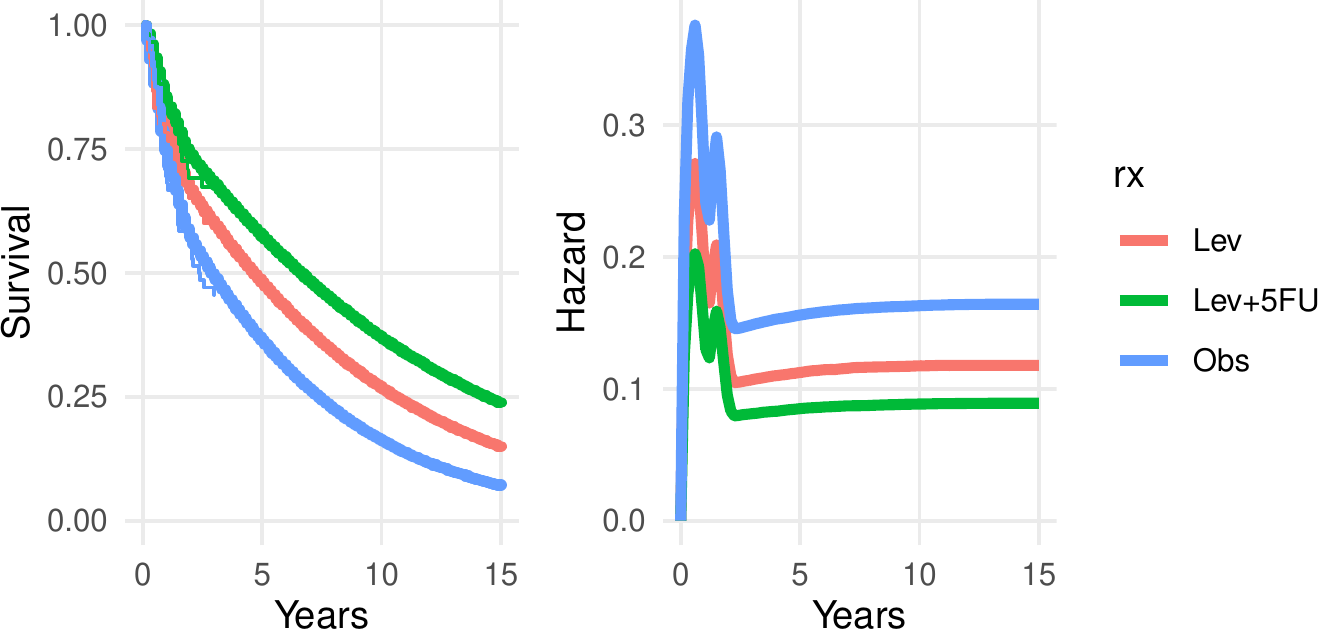} \end{center}

Examples of more advanced usage, and thorough documentation for all functions and features, are available at \url{http://chjackson.github.io/survextrap}.

\end{document}